\documentclass[reprint, amsmath, amssymb, aps,twocolumn] {revtex4-1}
\usepackage[dvipdfmx]{graphicx}
\usepackage{dcolumn}
\usepackage{bm}
\usepackage{verbatim}
\usepackage{wrapfig}
\usepackage{ascmac}
\usepackage{makeidx}
\usepackage{mathrsfs}
\usepackage{braket}
\usepackage{color}
\usepackage{ulem}
\usepackage[dvipdfmx]{hyperref}
\newcommand{\CsCrF}{CsCrF$_{4}$}
\begin{document}
\preprint{APS/123-QED}
\title{Inelastic Neutron Scattering in Weakly Coupled Triangular Spin Tubes CsCrF$_4$}
\author{Hodaka Kikuchi}
\affiliation{ Institute for Solid State Physics, University of Tokyo, Kashiwa, Chiba 277-8581, Japan}
\author{Shinichiro Asai}
\affiliation{ Institute for Solid State Physics, University of Tokyo, Kashiwa, Chiba 277-8581, Japan}
\author{Hirotaka Manaka}
\affiliation{ Graduate School of Science and Engineering, Kagoshima University, Korimoto, Kagoshima 890-0065, Japan}
\author{Masato Hagihala}
\affiliation{ Neutron Science Division, Institute of Materials Structure Science, High Energy Accelerator Research Organization, Tsukuba, Ibaraki 305-0801, Japan}
\author{Shinichi Itoh}
\affiliation{ Neutron Science Division, Institute of Materials Structure Science, High Energy Accelerator Research Organization, Tsukuba, Ibaraki 305-0801, Japan}
\author{Takatsugu Masuda}
\affiliation{ Institute for Solid State Physics, University of Tokyo, Kashiwa, Chiba 277-8581, Japan}
\affiliation{Institute of Materials Structure Science, High Energy Accelerator Research Organization, Ibaraki 305-0801, Japan}
\affiliation{Trans-scale Quantum Science Institute, The University of Tokyo, Tokyo 113-0033, Japan}



\date{\today}

\begin{abstract}
We performed inelastic neutron scattering (INS) experiments to measure spin dynamics on a polycrystalline sample of a spin tube candidate CsCrF$_{4}$. 
The compound exhibits a successive phase transition from a paramagnetic phase through an intermediate temperature (IT) phase of a 120$^{\circ}$ structure to a low temperature (LT) phase of another 120$^{\circ}$ structure. 
Elaborate comparison between observed and calculated neutron spectra in LT phase reveals that the spin Hamiltonian is identified as antiferromagnetic spin tubes including perturbative terms of intertube interaction, Dzyaloshinskii-Moriya interaction, and single ion anisotropy. 
A phase diagram for the ground state is classically calculated. 
A set of parameters in the spin Hamiltonian obtained from the INS spectra measured in LT phase is quite close to a boundary to the phase of the 120$^{\circ}$ structure of IT phase. 
The INS spectra measured in IT phase is, surprisingly, the same as those in LT phase in the level of powder averaged spectra, even though the magnetic structures in IT and LT phases are different. 
Identical dynamical structures compatible with two different static structures are observed. 
No difference in the observed spectra indicates no change of the spin Hamiltonian with the temperature, suggesting that the origin of the successive phase transition being order-by-disorder mechanism. 
\end{abstract}

\pacs{Valid PACS appear here}
\maketitle


\section{\label{sec:level1}Introduction}

Geometrically frustrated magnets have attracted great interest because of non-trivial magnetic states induced at low temperatures. 
The magnetic states of lattices including triangular basic unit are incompatible with N{\'e}el order, and they are macroscopically degenerated~\cite{GSdege}. 
An example of a one-dimensional frustrated system is a triangular spin tube where antiferromagnetic (AFM) spins on triangles are arrayed in one dimension, which has been theoretically studied extensively. 
In a regular triangular spin tube with Heisenberg spin, the ground state is the dimerized non-magnetic state having the unit of the two-site rung singlet~\cite{spintube1}. 
The spin correlation is exponentially decayed, and the excited state is separated by a finite spin gap~\cite{Wang2001, Luscher2004, Nishimoto2008, Schmidt2010, Nishimoto2011, Lajko2012}. 
In an asymmetric triangular spin tube where $Z_2$ symmetry is broken, the spin gap is suppressed, and 
Tomonaga$–$Luttinger liquid (TLL) with vector chiral order is predicted~\cite{tll1,tll2}. 
In contrast with the accumulative theoretical studies, experimental study has been limited to an $S$ = 3/2 spin tube candidate CsCrF$_4$~\cite{Manaka09,CsCrF1,CsCrF4,CsCrF0,CsCrF2,Hayashida20} for lack of a model material. 

\begin{figure}
\includegraphics[width=8.5cm]{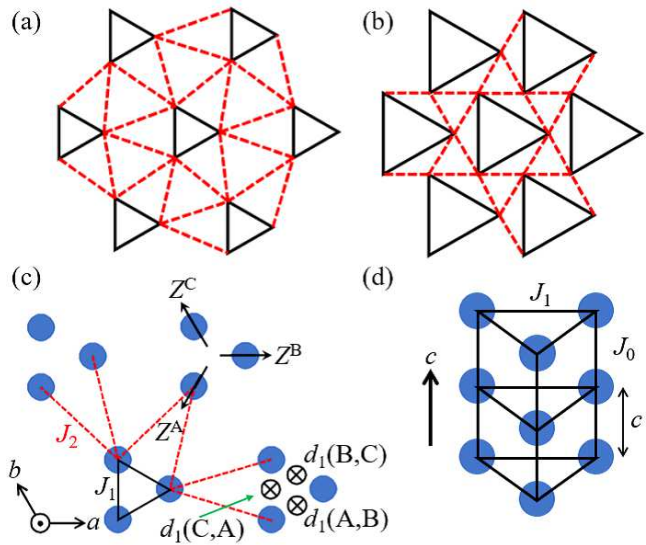}
\caption{(a) An example of geometry of coupled spin tubes viewed from the tube direction. 
(b) Schematic description of Kagome-Triangular lattice. (c) Crystal structure of CsCrF$_{4}$ projected on the crystallographic $ab$-plane. Blue circles indicate Cr$^{3+}$ ion. 
$Z^{\alpha}(\alpha$= A, B, C) is the $Z$ axis locally defined on $\alpha$ site of Cr$^{3+}$ ion, and ${\bm d_{1}}(\alpha,\beta) (\alpha,\beta$=A,B,C) is DM vector. 
(d) Perspective view of the spin tube of CsCrF$_{4}$. }
\label{CsCrF4structure}
\end{figure}
In a real compound, exchange interactions between triangular spin tubes are not negligible, and, hence, the geometry of the interactions between the tubes is important for determination of the ground state. 
An example of the geometry is shown in Fig.~\ref{CsCrF4structure}(a), where the regular triangles form two-dimensional triangular lattice. 
The lattice is equivalent to Kagome lattice having second neighbor interaction, which is called Kagome-Triangular (KT) lattice, as shown in Fig.~\ref{CsCrF4structure}(b). 
The ground state of a Heisenberg classical Kagome antiferromagnet is magnetically ordered state, 120$^{\circ}$ structure with a macroscopic degeneracy which is vulnerable to a fluctuation. 
Coplanar 120$^{\circ}$ structures having large entropy, where spins are confined in a plane, are selected at finite temperatures by order by disorder mechanism~\cite{obd1,obd2,obd3}. 
Experimentally, various types of 120$^{\circ}$ structures with ${\bm Q}$ = 0 were reported: the structure with positive chirality selected by Dzyaloshinskii-Moriya interaction in Fe- and Cr jarosite~\cite{jaro1,jaro2,jaro3,jaro4,jaro5}, that with negative chirality in semimetals Mn$_3$Sn and Mn$_3$Ge~\cite{Nagamiya82}, and tail-chase structure selected by magnetic dipole interaction in Mn$^{2+}$ fluoride~\cite{Hayashida18,Hayashida20b}. 
Introduction of the second neighbor ferromagnetic interaction in Heisenberg KT lattice leads to a cuboc structure, 
a noncoplanar multi-$Q$ structure having a 12-sublattice with the spins directing along the 12-middle points of a cube, and that of the second-neighbor antiferromagnetic interaction leads to $\sqrt{3} \times \sqrt{3}$ structure~\cite{cuboc1,KTL2,Seki15}. 

CsCrF$_{4}$ is a candidate of the triangular spin tube where intertube and rung couplings form KT lattice~\cite{Manaka09,CsCrF1,CsCrF4,CsCrF0,CsCrF2,Hayashida20}. 
The crystal structure of CsCrF$_{4}$, where Cr$^{3+}$ ions carrying spin $S$ = 3/2 are displayed and Cs$^{+}$ and F$^{-}$ ions are omitted, is shown in Figs. \ref{CsCrF4structure}(c) and \ref{CsCrF4structure}(d). 
Speculated from the bond lengths, exchange interactions along the leg $J_0$ and the rung $J_1$ in the spin tube would be dominant, and one between the spin tubes $J_ 2$ would be weak.
Cr$^{3+}$ network in the $ab$ plane shown in Fig.~\ref{CsCrF4structure}(a) is equivalent to KT lattice in Fig.~\ref{CsCrF4structure}(b), where the nearest neighbor Kagome interaction corresponds to $J_2$ and the second neighbor one corresponds to $J_1$. 
The magnetic susceptibility of this compound has a broad maximum at $T$ $\simeq$ 60 K, and hysteresis is observed below $T$ = 4 K~\cite{Manaka09}. 
Neutron diffraction scattering experiments probed a successive phase transition with the critical temperatures $T_{\rm N1} = 2.8$ K and $T_{\rm N2} = 3.5$ K~\cite{CsCrF2}. 
The magnetic structure of the intermediate temperature (IT) phase at $T_{\rm N1} \le T \le T_{\rm N2}$ is a $120^{\circ}$ structure with ${\bm q}_{\rm m} = (1/3, 1/3, 1/2)$, and that of the low temperature (LT) phase at $T \le T_{\rm N1}$ is a $120^{\circ}$ structure with ${\bm q}_{\rm m} = (1/2, 0, 1/2)$. 
Here ${\bm q}_{\rm m}$ is a propagation vector. 
The structure of LT phase is different from theoretical prediction of cuboc or $\sqrt{3} \times \sqrt{3}$ structure in Heisenberg KT lattice~\cite{cuboc1,KTL2,Seki15}. 
It was proposed that small perturbations including intertube interaction, DM interaction and single-ion anisotropy selected the observed magnetic structure in LT phase. Nevertheless, the spin Hamiltonian has not been identified yet.

In this study, we have performed inelastic neutron scattering (INS) experiments to measure the spin dynamics on a polycrystalline sample of \CsCrF . 
The spin Hamiltonian was identified as weakly coupled antiferromagnetic spin tubes by KT geometry. 
An intertube interaction, a single-ion anisotropy and DM interaction play important role in the selection of the ground state. 
Set of parameters in the spin Hamiltonian suggested that the compound is located in the phase of ${\bm q}_{\rm m} = (1/2, 0, 1/2)$, and is close to the boundary to the phase of ${\bm q}_{\rm m} = (1/3, 1/3, 1/2)$. 
The spectra both in IT and LT phases were measured, and they were qualitatively the same. 
Identical dynamical structures compatible with different static structures were observed for the first time. 
Order-by-disorder mechanism was suggested for the origin of the successive phase transition.

\section{\label{sec:level1}Experimental details}

To collect INS spectra in wide momentum ($Q$) -- energy ($\hbar \omega$) space, INS experiments were carried out by High Resolution Chopper spectrometer (HRC)~\cite{ITOH201190} installed at BL12 in J-PARC/MLF. 
A polycrystalline sample of \CsCrF\ with a mass of 5.07 g prepared by a solid state reaction method was used \cite{Manaka09,CsCrF1}. 
$^3$He cryostat was used to achieve low temperatures. 
Frequency of Fermi chopper was 100 Hz, and incident neutron energies ($E_{\rm i}$) were 3.05, 6.10 and 15.3 meV. 
The data reduction was performed by HANA software~\cite{Kawana_2018}. 

To collect temperature dependence of low-energy excitation, INS experiments were carried out by High Energy Resolution cold neutron triple-axis spectrometer (HER) installed at C11 beam port in JRR-3.
A polycrystalline sample with a mass of 2.69 g was used. 
ORANGE-type cryostat was used. 
The collimator setup was guide-open-radial collimator-open. 
A Be/PG filter for elimination of the second order harmonics of neutron, which automatically switches at $E_{\rm i} = 5$ meV, was set in front of the sample. 
Be filter was used for $E_{\rm i} \le$ 5 meV. 
A tunable PG filter~\cite{Freund85,Vorderwisch99} was used for $E_{\rm i} >$ 5 meV. 
Final neutron energy was fixed at $E_{\rm f} = 3.64$ meV.

\section{\label{sec:level1}Experimental results}

\begin{figure}
\includegraphics[width=8.5cm]{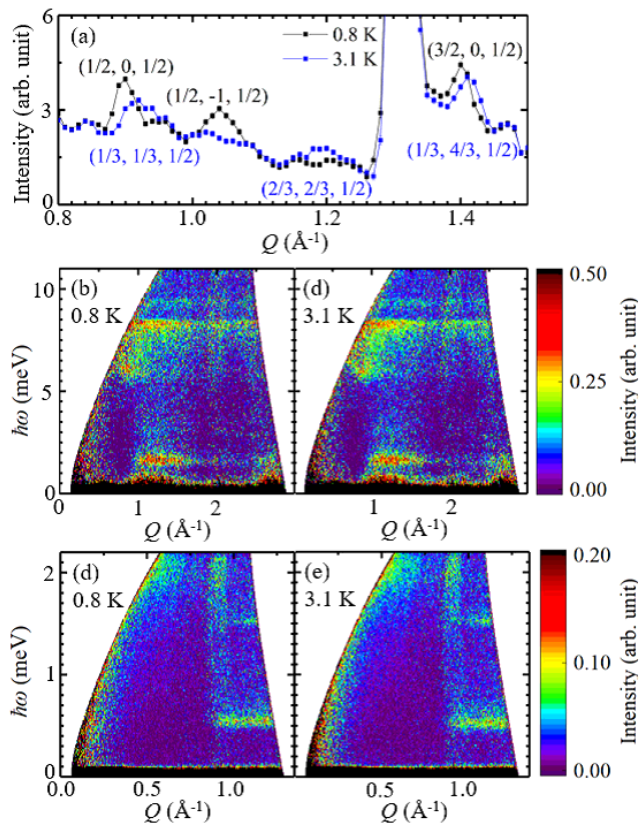}
\caption{INS spectra measured by HRC spectrometer. 
(a) Elastic part of INS spectra
measured at $E_{\rm i}=6$ meV, where integration range is $-0.1 \le E \le 0.1$ meV. 
Black and blue symbols indicate 0.8 K and 3.1 K, respectively. 
INS spectra measured with $E_{\rm i} = 15.3$meV at 0.8 K for (b) and 3.1 K for (c). INS spectra measured with $E_{\rm i} = 3.05$meV at 0.8 K for (d) and 3.1 K for (e).}
\label{exp_HRC}
\end{figure}
Elastic part of measured INS spectra at 0.8 K in LT phase and at 3.1 K in IT phase using HRC spectrometer are shown in Fig.~\ref{exp_HRC}(a). 
Magnetic peaks with ${\bm q}_{\rm m} = (1/2, 0, 1/2)$ are observed at 0.8 K, and those with ${\bm q}_{\rm m} = (1/3, 1/3, 1/2)$ are observed at 3.1 K. 
These results are consistent with previous neutron diffraction~\cite{CsCrF2}.

Figures \ref{exp_HRC}(b) and \ref{exp_HRC}(d) show INS spectra in LT phase measured at $E_{i} = 15.3$ and 3.05 meV, respectively. 
Magnetic excitations up to 10 meV are observed, with flat intensities at 0.5 meV and 1.5 meV, and a continuous intensity at 6 $\sim$ 8 meV. 
The flat intensities modulate with periodicity of about $1.6~{\rm \AA}^{-1}$, which coincides the reciprocal lattice constant in the $c^{\ast}$ direction. 
This fact implies that the system is quasi-one-dimensional with strong interaction in the $c$ direction. Furthermore, absence of intensity at $Q$ smaller than 0.8 $\rm\AA^{-1}$, which corresponds to $c^{\ast}$/2, indicates that the main interaction is antiferromagnetic rather than ferromagnetic. 
Note that Cr-Cr spacing along the $c$ axis is the lattice constant $c$ as shown in Fig.~\ref{CsCrF4structure}(d). 
Figures \ref{exp_HRC}(c) and \ref{exp_HRC}(e) show INS spectra in IT phase measured at $E_{i} = 15.3$ and 3.05 meV, respectively. 
Surprisingly, no qualitative difference is found between the spectra in LT and IT phases despite the change in the magnetic propagation vectors.

\begin{figure}
\includegraphics[width=8.4cm]{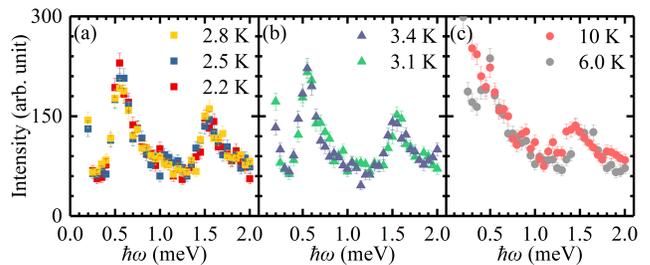}
\caption{Constant $Q$ scans at $Q= 1.05~\rm\AA^{-1}$ collected in LT phase in (a), IT phase in (b), and paramagnetic phase in (c) measured by HER spectrometer.}
\label{exp_HER}
\end{figure}

Temperature dependence of constant $Q$ scans measured at $Q = 1.05~\rm\AA^{-1}$ using HER spectrometer is shown in Figs.~\ref{exp_HER}(a)--\ref{exp_HER}(c). 
The square, triangle, and circle symbols indicate the data in LT, IT, and paramagnetic phases, respectively. 
Well-defined excitations are observed at 0.5 meV and 1.5 meV both in LT and IT phases, and the spectra are qualitatively the same. 
The results are consistent with those measured using HRC. 
At $T = 6$ K which is higher than $T_{\rm N2}$, the inelastic excitations at 0.5 meV and 1.5 meV are still observed. 
At $T = 10$ K, paramagnetic excitation is enhanced and the peaks are smeared. 

\section{\label{sec:level1}Analysis}

INS cross section of \CsCrF\ was calculated based on linear spin-wave theory (LSWT) using SpinW package~\cite{SpinW}. 
Analytic approximation was adopted for the magnetic form factor of Cr$^{3+}$ ions~\cite{Dianoux2003NeutronDB}.
The following Hamiltonian was used according to previous study~\cite{CsCrF2}:
\begin{eqnarray}
{\mathcal H}&=&\sum_{\bm{l}_{\alpha},\bm{l}_{\beta},\bm{l'}_{\beta}}\{J_{0}\bm{S}(\bm{l}_{\alpha})\cdot {\bm S}(\bm{l}_{\alpha}+\bm{c})+J_{1}\bm{S}(\bm{l}_{\alpha})\cdot {\bm S}(\bm{l}_{\beta}) \nonumber \\
&+& J_{2}\bm{S}(\bm{l}_{\alpha})\cdot {\bm S}(\bm{l'}_{\beta})+\bm{d}_{1(\alpha,\beta)}\cdot \bm{S}(\bm{l}_{\alpha})\times\bm{S}(\bm{l}_{\beta}) \nonumber \\
&+& D(\bm{S}^{{Z}^{\alpha}}(\bm{l}_{\alpha}))^{2}\},
\label{eq1}
\end{eqnarray}

where $J_0$ and $J_1$ are leg and rung interactions in the spin tube in Fig.~\ref{CsCrF4structure}(d), and $J_2$ is 
an intertube interaction in Fig.~\ref{CsCrF4structure}(c). 
$\bm{d}_{1(\alpha,\beta)}$ is DM vectors between $\alpha$ and $\beta$ sites of Cr$^{3+}$ ions in the triangle, $D$ is a single-ion anisotropy, and ${Z}^{\alpha}$ is the local $Z$ axis defined at the $\alpha$ site as shown in Fig.~\ref{CsCrF4structure}. 
$\bm{l}_{\alpha}$ is the position vector of $\alpha$ site in the lattice ${\bm l}$, and ${\bm c}$ is the unit vector of the crystal lattice along the $c$ axis. 
The sum is taken for $\bm{l}_{\alpha}$ ,$\bm{l}_{\beta}$, and $\bm{l'}_{\beta}$ all over the crystal, where $\alpha \ne \beta$ and ${\bm l} \ne {\bm l'}$. 

\begin{figure}
\includegraphics[width=8.5cm]{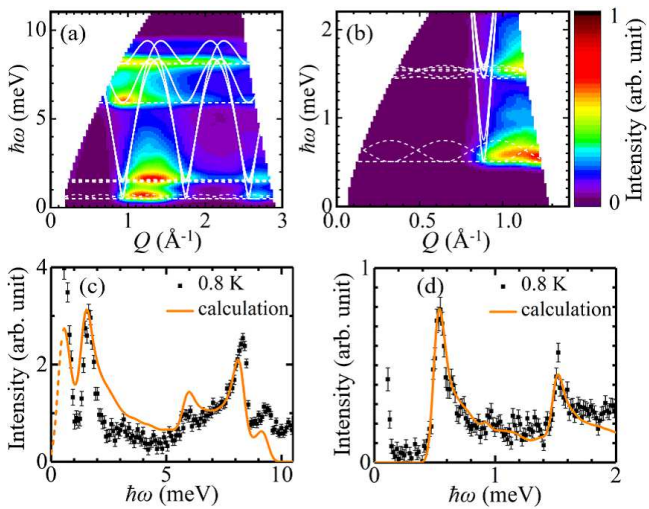}
\label{fig:cal}
\caption{One-dimensional cuts of measured and calculated INS spectra in wide (a) and narrow (b) energy ranges. Symbols and solid curves indicate experiment and calculation, respectively. INS spectra calculated by linear spin wave theory in wide (c) and narrow (d) energy ranges. 
Solid and dashed curves in (c) are dispersion curves calculated on single crystal in the $c^{*}$ and $a^{*}$ directions, respectively. 
}
\label{cal}
\end{figure}

One-dimensional (1D) cuts of measured and calculated INS spectra are shown in Figs.~\ref{cal}(a) for wide and \ref{cal}(b) for narrow energy ranges. 
The measured spectrum in the former is cut from Fig.~\ref{exp_HRC}(b) with the integration range of $1.2~{\rm \AA}^{-1}\le Q \le 1.5~{\rm \AA}^{-1}$, and that in the latter is cut from Fig.~\ref{exp_HRC}(d) with the range of $1.0~{\rm \AA}^{-1}\le Q \le 1.1~{\rm \AA}^{-1}$. 
Parameters $J_0$, $J_1$, $\bm{d}_{1(\alpha,\beta)}$, and $D$ were determined so that the peak energies of the calculation reproduce those of the experiment. 
Errors were determined by the intervals of the parameters in the search. 
The obtained parameters are summarized in Table~\ref{tab:table1}. 
The digits in the parenthesis are the errors. 
Estimate of $J_2$ will be explained later. 
The calculated spectra of the polycrystalline sample using Eq.~(\ref{eq1}) and parameters in Table \ref{tab:table1} are shown in Fig. \ref{cal}(c) for wide and Fig.~\ref{cal}(d) for narrow energy ranges. 
The instrumental resolutions for $Q$ and $\hbar \omega$ were convoluted in Figs. \ref{cal}(c), and \ref{cal}(d). 
The calculation reasonably reproduces the measured spectra in Figs.~\ref{exp_HRC}(b) and \ref{exp_HRC}(d). 
Since the leg ($J_0$) and rung ($J_1$) interactions are dominant and the intertube ($J_2$) one is small, the spin Hamiltonian is identified as antiferromagnetic spin tubes which are weakly coupled by KT geometry. 
The white dashed and solid curves in Fig. \ref{cal}(c) show the dispersion curves of the single crystal in the $c^{*}$ and $a^{*}$ directions, respectively.
The energies of the flat intensities at 0.5, 1.5, 6.0, and 8.0 meV in the polycrystalline spectrum are identical to the dispersion curves of the single crystal in the $a^{*}$ direction. 
They are almost independent of $Q$ because of small $J_2$. 
The $Q$ dependence is rather enhanced at 0.5 meV, but it is not detected experimentally. 
The instrumental resolution was, thus, used to estimate the lower limit of $J_2 ( < 0)$, where $J_2$ was taken ferromagnetic to ensure the spin structure reported previously~\cite{CsCrF2}. 
The lower limit, which was the maximum of the absolute value in this case, was estimated as $-7~\mu$eV. 
The round off of the half of the lower limit, $-4~\mu$eV, was determined as $J_2$.
By comparison between the polycrystalline and single crystalline spectra, it is found that well defined flat intensities in an INS polycrystalline spectrum are ascribed to one dimensionality of a spin system. 
In contrast, the dispersion in the $c^{*}$ direction induces intensity spread in wide $Q$ - $\hbar \omega$ 
range. 
The ratio of $J_{0}$ and $J_{1}$ was reported to be about $1/2$ in a first-principles calculation~\cite{fpc}. 
The ratios estimated from the present experiment is in good agreement with the calculation. 

\begin{table}
\caption{
Parameters of the spin Hamiltonian. 
$J_0$, $J_1$, $d_1$, and $D$ are determined by the comparison between calculation and experiment in LT phase. 
$J_2$ is determined from the lower limit of $J_2$ estimated by the instrumental energy resolution (see text). 
}
\begin{ruledtabular}
\begin{tabular}{ccccc}
$J_0$ (meV) & $J_1$ (meV) & $J_2$ ($\mu$eV) & $d_1$ ($\mu$eV) & $D$ ($\mu$eV)\\
\hline
2.35 (5) & 1.00 (5) & $-4(3)$ & $-47 (2)$ & $-4.6 (2)$
\end{tabular}
\end{ruledtabular}
\label{tab:table1}
\end{table}

\section{\label{sec:level1}Discussion}

\begin{figure}
\includegraphics[width=7cm]{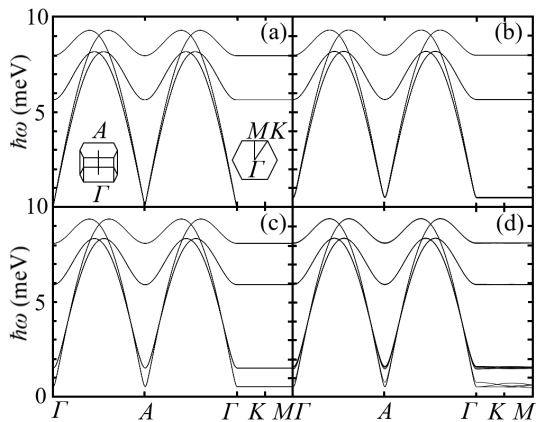}
\label{fig:cal}
\caption{Dispersion curves of single crystalline sample of CsCrF$_{4}$ calculated by four sets of spin parameters described in Table~\ref{table1a}: (a) case 1, (b) case 2, (c) case 3, and (d) case 4.}
\label{dispersion}
\end{figure}

To understand influence of each parameter in the spin Hamiltonian on INS spectrum, dispersion curves among reciprocal lattice points with high symmetry were calculated in Figs.~\ref{dispersion}(a)--\ref{dispersion}(d) for four cases, where the parameters are shown in Table~\ref{table1a}. 
In case 1, which corresponds to an isolated spin tube, the excitation is dispersive along the $c^{*}$ axis (from $\Gamma$ to $A$) and dispersionless perpendicular to the $c^{*}$ axis (from $\Gamma$ through $K$ to $M$). 
Periodicity along the $c^{*}$ axis is halved of the crystal lattice because of the antiferromagnetic leg interaction $J_0$. 
Our analysis suggests that two modes appear at the energy of $\sqrt{27J_{0}J_{1}/2}$ and $\sqrt{27J_{0}J_{1}}$ at the $\Gamma$ and $A$ points. 
This means that the dispersive modes in high energy region of 5.5 meV $\le \hbar \omega \le$ 9 meV are 
from the rung interaction $J_1$.

\begin{table}
\caption{
Parameters of the spin Hamiltonian for dispersion curves of single crystalline sample used in Figs.~\ref{dispersion}(a)--\ref{dispersion}(d).
}
\begin{ruledtabular}
\begin{tabular}{cccccc}
&$J_0$ (meV) & $J_1$ (meV) & $J_2$ ($\mu$eV) & $d_1$ ($\mu$eV) & $D$ ($\mu$eV)\\
\hline
Case 1 &2.35 & 1.00 & 0 & 0 & 0 \\
Case 2 &2.35 & 1.00 & 0 & 0 & $-4.6$ \\
Case 3 &2.35 & 1.00 & 0 & $-47$ & $-4.6$ \\
Case 4 &2.35 & 1.00 & $-4$ & $-47$ & $-4.6$
\end{tabular}
\end{ruledtabular}
\label{table1a}
\end{table}
In case 2, as shown in Fig. \ref{dispersion}(b), the single ion anisotropy, $D$, breaks the rotational symmetry in the $ab$ plane, lifts the degeneracy of the ground state, and induces an energy gap of 0.5 meV at $(h, k, 1/2)$ including $\Gamma$, $A$, $K$, and $M$ points. 
In case 3, as shown in Fig. \ref{dispersion}(c), DM interaction, $d_1$, lift the first excited state, and another mode appears at 1.5 meV. 
In case 4, as shown in Fig. \ref{dispersion}(d), the intertube interaction, $J_{2}$, makes the flat mode dispersive in the $ab$ plane. 
The dispersion is larger at lower energies, but still within the energy resolution. 
The results indicate that the effect of intertube interaction is negligibly small in \CsCrF\ even though it is important for the determination of the ground state. 
Dynamics of this compound is, thus, dominated by the isotropic spin-tube, and the effect of small perturbations is detected as the energy gaps at the low energies. 

In coplanar 120$^{\circ}$ structure in a Kagome lattice, certain spins surrounded by spins pointing to a same direction, such as six spins of a hexagon or spins on a certain line, can be locally tilted from the plane without energy cost, i.e., with keeping the angle among the adjacent spins at $120^{\circ}$. 
Such a mode is called zero-energy mode~\cite{ZEM1,ZEM2}. 
The mode is lifted by DM and/or dipole interactions, and it is experimentally observed as flat excitations~\cite{ZEM3,ZEM4,ZEM2,Hayashida20b}. 
The flat modes in CsCrF$_{4}$ is, however, different from the zero-energy mode because it appears as the result of the lift of the degeneracy of the ground state in the quasi-isolated triangular spin system by the single ion anisotropy and DM interaction.

\begin{figure}
\includegraphics[width=8.5cm]{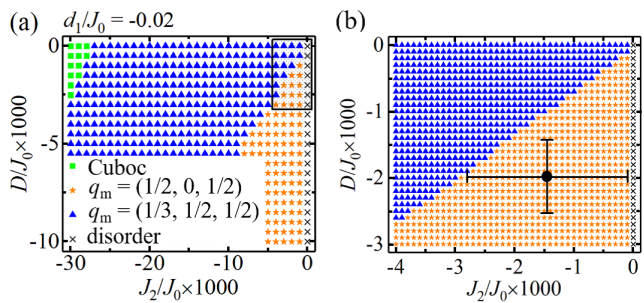}
\caption{A phase diagram of CsCrF$_{4}$ calculated by Luttinger-Tisza method. (a) $D-J_{2}$ phase diagram with $J_{0} = 1,~J_1 = 0.43,~{\rm and}~d_1 = -0.02$. (b) Enlarged phase diagram of the region surrounded by a solid rectangular in (a). }
\label{PD}
\end{figure}
A classical phase diagram of the ground state using Luttinger-Tisza method~\cite{ltm1,ltm2} was calculated using Eq.~(\ref{eq1}) as shown in Fig.~\ref{PD}(a), where $J_{0}$ was set to 1. 
Cuboc structure, 120$^{\circ}$ structure with ${\bm q}_{\rm m} = (1/2, 0, 1/2)$, and $\sqrt{3} \times \sqrt{3}$ structure with ${\bm q}_{\rm m} = (1/3, 1/3, 1/2)$ exist. 
Disordered state of isolated spin tubes at $J_2$ = 0 is indicated by crosses. 
The region surrounded by a solid rectangular in Fig.~\ref{PD}(a) is enlarged in Fig.~\ref{PD}(b). 
A filled circle is the position of the parameters obtained in the present study. 
\CsCrF\ locates in the phase of ${\bm q}_{\rm m} = (1/2, 0, 1/2)$, which is consistent with the magnetic structure in LT phase~\cite{CsCrF2}.
It is quite close to the phase boundary to ${\bm q}_{\rm m} = (1/3, 1/3, 1/2)$ phase. 

Let us discuss origin of the successive phase transition. 
IT phase with ${\bm q}_{\rm m} = (1/3, 1/3, 1/2)$ structure can be stable if small lattice distortion suppresses $D$ or enhances $J_2$ with the temperature. 
The INS spectra in IT and LT phases were, however, the same, and the change of the parameters would not be the origin. 
For confirmation, we calculated the INS spectra of IT phase using parameters located in the phase of ${\bm q_{\rm m}}=(1/3,1/3,1/2)$ near the boundary to the phase of ${\bm q}_{\rm m} = (1/2, 0, 1/2)$. 
The calculated spectra were totally different from the observed one; small change in the parameters made drastic change in the spectrum no matter how the change was small, which is described in Appendix A. 
Therefore, the origin of the successive transition in CsCrF$_{4}$ is not change of the parameters in the spin Hamiltonian, but realization of a state having large entropy by order by disorder mechanism.

There are few previous studies on the comparison of INS spectra in IT and LT phases in spite of existence of numerous compounds which undergo a successive phase transition: 
$\rm{LiNiPO_{4}}$ \cite{LNPO} and $\rm{Li_{2}NiW_{2}O_{8}}$ \cite{LNWO} which change from commensurate structure in IT phase to incommensurate structure in LT phase, $\rm{CsCoCl_{3}}$ \cite{CCC} and $\rm{CuFeO_{2}}$ \cite{Mekata93,Mitsuda98} which change from partial order to full order, and $\rm{CsMnI_{3}}$ \cite{CMI} 
which changes from a collinear to $120^{\circ}$ structure with keeping the same propagation vector. 
Among them, $\rm{CsMnI_{3}}$ is the only compound, for which spectra both in IT and LT phases were measured. 
The spectra, including dispersion curves and intensities, in IT and LT phases were clearly different, even though the propagation vectors were identical. 
In contrast in \CsCrF , the spectra in IT and LT phases are identical, while the propagation vectors are different. 
The present study is the first demonstration that dynamical structure does not change even though static structure changes in a compound exhibiting a successive transition. 

The ground state of spin $S = 3/2$ triangular spin tube is a valence bond solid state with dimerization, and an energy gap opens between the ground and excited states~\cite{Nishimoto2011}. 
Such a gap has been observed in spin gap systems including spin $S$ = 1 Haldane chain, spin $S$ = 1/2 spin-Peierls, spin ladder, etc., by INS experiment. 
The spin gap was observed even in the compound having magnetically ordered state induced by interchain interaction or impurity such as Haldane chain CsNiCl$_3$~\cite{Buyers86}, impurity-doped spin-Peierls compound CuGeO$_3$~\cite{Martin97}, and impurity-doped Haldane chain PbNi$_2$V$_2$O$_8$~\cite{Uchiyama99,Zheludev00}. 
In the case of the spin $S = 3/2$ spin tube compound CsCrF$_{4}$ in the present study, the gap is estimated to be about 0.006 meV ($= 0.0025J_{\rm leg}$)~\cite{Nishimoto2011}. 
This is much smaller than the instrumental resolution, and it was not observed in the present study.

\section{\label{sec:level1}Summary}

In summary, the spin Hamiltonian of \CsCrF\ was identified as weakly coupled antiferromagnetic spin tubes by INS experiments. 
A set of obtained parameters from the experiment in LT phase was quite close to the phase boundary between the 120$^{\circ}$ structure with ${\bm q_{\rm m}} = (1/2, 0, 1/2)$ and one with ${\bm q_{\rm m}} = (1/3, 1/3, 1/2).$ 
The observed INS spectra in IT and LT phases were qualitatively the same even though the magnetic structures were different. 
The origin of the successive phase transition turned out to be order-by-disorder mechanism. 
Observation of identical dynamical structures compatible with different static structures in the present study is an important issue. 
Theoretical examination of the spin dynamics at finite temperatures in a frustrated system exhibiting a successive phase transition is awaited. 



\begin{acknowledgements}
We are grateful to D. Kawana, T. Asami, and R. Sugiura for supporting us in the neutron scattering experiment at HRC and HER. 
The neutron experiment using HRC spectrometer at the Materials and Life Science Experimental Facility of the J-PARC was performed under a user program (Proposal No. 2019S01). 
The neutron experiment using HER at JRR-3 was carried out by the joint research in the Institute for Solid State Physics, the University of Tokyo (Proposal No. 21403). 
H. Kikuchi was supported by Support for Pioneering Research Initiated by Next Generation (SPRING) of Japan Science and Technology Agency (JST) . 
This project was supported by JSPS KAKENHI Grant Numbers 19KK0069, 20K20896 and 21H04441. 
\end{acknowledgements}

\appendix

\section{Calculated INS Spectra for Spin Structure in IT Phase}

\begin{figure}
\includegraphics[width=9cm]{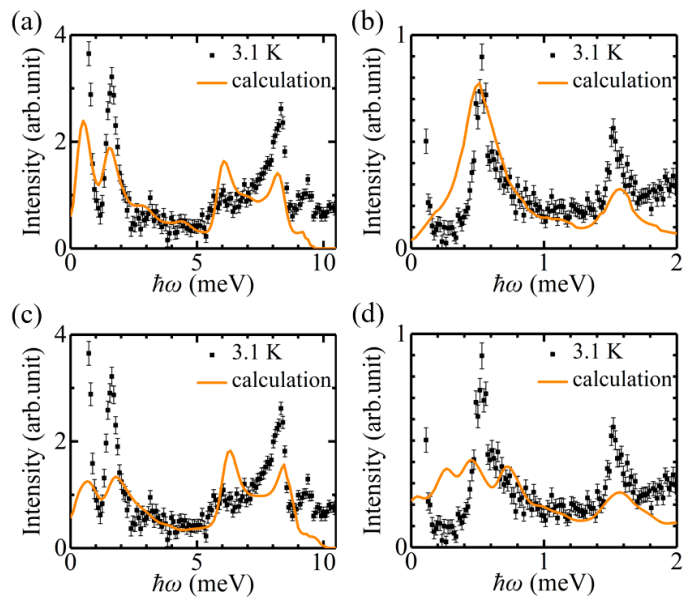}
\caption{
1D cuts of calculated INS spectra (orange curves) by using the spin parameters summarized in Table~\ref{table2} 
and those of experiment (symbols) measured at 3.1 K in IT phase. 
The spectra of case A1 in (a) for wide and (b) for narrow energy ranges. 
The spectra of case A2 in (c) for wide and (d) for narrow energy ranges. 
}
\label{sub_fig2}
\end{figure}

INS spectra from the ground state with the structure of ${\bm q_{\rm m}}=(1/3,1/3,1/2)$ were calculated 
for two set of parameters, case 1 and 2, in Table~\ref{table2}.
In case 1, $D$ was fixed to 0, and $J_{0}$, $J_{1}$, $J_{2}$, and $d_1$ were determined so that the calculated energy of the excitations in IT phase in 1D cuts coincides with the observed one, as were determined in LT phase. 
In case 2, $J_2$ was fixed to $-8~\mu$eV and $D$ was fixed to $-4.6~\mu$eV, and other parameters were determined in the same method as case 1. 
The calculated 1D cuts using the best parameters in case 1 in Table \ref{table2} are shown by orange curves in Fig.~\ref{sub_fig2}(a) for wide and Fig.~\ref{sub_fig2}(b) for narrow energy ranges. 
The excitation with the peak energy of 0.5 meV can be produced by $J_2$ instead of $D$, 
and, however, the peak width is wider than the energy resolution in Fig.~\ref{sub_fig2}(b). 
The intensities of the peaks, particularly at 6 and 8 meV, in Fig.~\ref{sub_fig2}(a) are not reproduced.
The calculated 1D cuts using the best parameters in case 2 in Table \ref{table2} are shown by orange curves in Fig.~\ref{sub_fig2}(c) for wide and in Fig.~\ref{sub_fig2}(d) for narrow energy ranges. 
The low-energy mode at $\hbar \omega \sim 0.5$ meV is lifted both by $D$ and $J_2$, leading to a complex profile. 
The measured spectrum is, thus, not reproduced by the calculation. 
The change of the parameters leads to drastic change of INS spectrum, even though the change is small. 

\begin{table}
\caption{
Parameters of the spin Hamiltonian for the calculation of the INS spectra in IT phase. 
}
\begin{ruledtabular}
\begin{tabular}{cccccc}
&$J_0$ (meV) & $J_1$ (meV) & $J_2$ ($\mu$eV) & $d_1$ ($\mu$eV) & $D$ ($\mu$eV)\\
\hline
Case A1 &2.55 (5) & 1.10 (5) & $-7$ (1) & $-51$ (1) & 0 \\
Case A2 &2.39 (5) & 1.02 (5) & $-8$ & $-55$ (1) & $-4.6$ 
\end{tabular}
\end{ruledtabular}
\label{table2}
\end{table}

\clearpage

\bibliography{INS_CsCrF4_Kikuchi_ver1}

\end{document}